\documentclass{ws-ijbcax}
\usepackage{graphicx}
\usepackage{epstopdf}

\makeatletter
  \newcounter{subeqncnt}
  \def\thesubeqncnt{\arabic{subeqncnt}}
  \def\subequations{\begingroup%
      \edef\@tempa{\theequation}%
     \let\c@equation\c@subeqncnt\c@subeqncnt\z@
      \edef\theequation{L\noexpand\thesubeqncnt}}
  
  \makeatother

\begin{document}

\catchline{}{}{}{}{} 

\markboth{M. Yabuki and T. Tsuchiya}{Double Precision Computation of the Logistic Map Depends on Computational Modes of the Floating-point Processing Unit}

\title{DOUBLE PRECISION COMPUTATION OF THE LOGISTIC MAP DEPENDS ON COMPUTATIONAL
 MODES OF THE FLOATING-POINT PROCESSING UNIT}

\author{MICHIRO YABUKI
}

\address{Department of Information Science, Meisei University, 2-1-1 Hodokubo\\
Hino-shi, Tokyo 191-8506, Japan\\
yabuki@is.meisei-u.ac.jp
}

\author{TAKASHI TSUCHIYA\footnote{Professor Emeritus}}
\address{Department of Information Science, Meisei University, 2-1-1 Hodokubo\\
Hino-shi, Tokyo 191-8506, Japan\\
tsuchiya@is.meisei-u.ac.jp}

\maketitle


\begin{abstract}
Today's most popular CPU can operate in two different computational modes for
double precision computations. This fact is not very widely recognized among
scientific computer users. The present paper reports the differences the modes 
bring about using the most thoroughly studied system in chaos theory,
the logistic map. Distinct virtual periods due to finite precision come about
depending on the computational modes for the parameter value corresponding to
fully developed chaos. For other chaotic regime various virtual periods emerge
depending on the computational modes and the mathematical expressions of the map.
Differences in the bifurcation diagrams due to the modes and the expressions are
surveyed exhaustively. A quantity to measure those differences is defined and 
calculated.
\end{abstract}


\section{Introduction}	\label{mytt:Intro}

Most numerical calculations are currently done in double precision(DP),
whose accuracy is approximately 16 decimal digits is well known among the
scientific computer users.  The following fact, however, might not be so well
recognized; today's most popular Central Processing Unit(CPU) by
Intel Corporation(as well as that by Advanced Micro Devices, Inc.) has the Floating-point Processing Unit(FPU)
which can operate in two different computational modes; one is the
DP formalized by IEEE 754 with 64 bits \cite{ieee7541} and the
other is called the extended precision mode utilizing 80 bits.
This difference in computational modes may
yield a slight discrepancy in the final results. As is easily expected,
the discrepancy is so small that it can be ignored as long as the absolute
tolerance is set to be larger than $10^{-16}$ in usual
calculations where convergent results are assumed.

However, for chaotic systems in which a small difference in the initial
states is extended to the size of the system itself \cite{chaosbook}, the discrepancy stated
above can possibly be manifested macroscopically. In this paper we report
that the differences in computational modes actually bring about the
differences in the final results using the simplest and most frequently
studied chaos generator, the logistic map. If we denote the system variable
by $x$, an adjustable external parameter by $a$ and count the number of
iterations by $n$, the logistic map is commonly expressed as
\begin{equation}
	x_{n+1}=ax_{n}(1-x_{n}).	\label{mytt:eq1}
\end{equation}
It is readily seen that the right-hand-side of Eq.(\ref{mytt:eq1}) can
be expressed in several different forms, all of which are, of course,
mathematically equivalent. We show, in this report, that there are cases
the different expressions of the logistic map in computer programs give
different macroscopic outputs depending on the two computational modes of FPU. 

Periodic orbits should not be observed when a system is in chaotic regime,
but it is already well known that the results of numerical calculations of
the logistic map (\ref{mytt:eq1}) in DP which has a finite precision eventually fall
into virtual periodic orbits, though their periods are very long \cite{perilen1}\cite{perilen2}\cite{perilen3}.
We here show further that the periodicity of the orbits thus produced
depends on the computational modes of FPU and on the expressions of the logistic 
map used in the computer program. 

The bifurcation diagram is commonly drawn in order to grasp dynamical
behaviors of the system under consideration for a wide range of external
parameter values. We show that the differences in the computational modes
of FPU and the expressions in the computer program bring about
differences in bifurcation diagrams. Furthermore, we propose a measure that can 
quantitatively distinguish the differences, and actually calculate the measure. 

Although the system treated in this paper as an example is only the logistic
map, our most general aim of this paper is to make the following proposal;
in reports of numerical investigations of chaotic systems explicit mentions
of mathematical expression used in the program and the computational mode
of FPU should be standardized in order to make the reports maximally
reproducible.

In the next section we explain about the two computational modes of FPU under
the condition that the precision is set at DP. In the third section we
present six possible expressions in computer programs for the logistic
map Eq.(\ref{mytt:eq1}). In section \ref{mytt:VP}, it is reported that periodicity that inevitably
emerges as an artifact when the logistic map in chaotic regime is computed
with DP. The point is that the period is strongly
dependent on the computational modes of FPU and mathematical expressions of
the map in the program used. In section \ref{mytt:DBD},  we exhaustively discuss
the difference between the bifurcation
diagrams when the bifurcation diagrams are computed using different modes of FPU and the map
expressions. In the final section we come back to
the proposal stated above.

\section{Computational Modes of FPU} \label{mytt:compmode}
IEEE 754-2008 formulates floating-point number for single precision as 32 bits, 
double precision(DP) as 64 bits and quadratic precision as 128 bits \cite{ieee7542}.  DP
is most prevalently used for numerical calculations today. In
DP, 1 bit is allocated for the sign, 11 bits for the exponent and 52 bits
(53 bits if one includes the so-called hidden bit which is a leading bit of normalized mantissa not actually stored in the datum)
for the mantissa. After each arithmetic operation, rounding is done.
The rounding modes are also formulated by IEEE 754-2008 as follows \cite{ieee7541},
namely, round-to-even, round-toward-zero, round-up and round-down. Since
most current FPU's default round-to-even rounding mode and most programmers
take that mode for granted, we restrict ourselves only to this rounding
mode.

Even if one restricts the rounding mode to round-to-even, there are two DP
computational modes in today's popular FPU's as mentioned in section \ref{mytt:Intro}. In one mode, 53 bits are
used for mantissa both for the left-hand-side and the right-hand-side of
each arithmetical expression in the program, while in the other mode only
the left-hand-side value has 53-bit-mantissa length and for the
right-hand-side value 64-bit-mantissa length is utilized. Since our main
interest in this paper is in the difference between these two modes, we
devise the following notation to distinguish the mode, namely,
\begin{equation}
	[\alpha : \beta]
\end{equation}
where $\alpha$ represents the left-hand-side mantissa bit-length and
$\beta$ the right-hand-side mantissa bit-length. Hence, we say there
are $[53:53]$ mode and $[53:64]$ mode for DP calculations. As far as
the present authors have surveyed, each of the currently popular compilers
defaults one of the two modes as:
\begin{tabbing}
\hspace{1cm}$[53:53]$ ~~~  \= GNU C Compiler (gcc) on FreeBSD(Version 8 and 9)\\
 \> gcc on MacOS X\\
 \> VC{\tt ++}2008 on Windows\\
 \> \\
\hspace{1cm}$[53:64]$ \> bcc 5.5 on Windows XP\\
 \> gcc on GNU/LINUX (Fedora 2.6.27.5).
\end{tabbing}

As we can easily imagine, there are situations that produce different
results depending on the mode we employ; for example (in this paper we
use C language to represent a part of a computer program)

\begin{verbatim}
        double x, y, z;
    
        x = 10.0;
        y = 2.718281810;
        z = x / (y * y);
        printf("%21.16e\n", z);
\end{verbatim}
yields 1.3533528507465618e+00 when the mode $[53:53]$ is used, and 
1.3533528507465620e+00 for the mode $[53:64]$ giving rise to the difference
$2.0\times 10^{-16}$ in decimal digit. For such cases as this example,
the difference can easily be ignored by setting the tolerance
larger than $10^{-15}$. But things are not that simple for chaotic systems
such as the logistic map in chaotic regime because each iterate of the map
can be different depending on its mathematical expression, as will be
presented in the following sections.

\section{Different Expressions for the Logistic Map}	\label{mytt:diffexp}

The logistic map which is most frequently written in the form
given in Eq.(\ref{mytt:eq1}) can be expressed in computer programs in the following
six distinct forms
\begin{subequations}
\begin{eqnarray}
	x_{n+1} & =(ax_{n})(1.0-x_{n})\\
	x_{n+1} & =a(x_{n}(1.0-x_{n}))\\
        x_{n+1} & =(a-ax_{n})x_{n}\\
        x_{n+1} & =a(x_{n}-x_{n}x_{n}) \\
        x_{n+1} & =ax_{n}-(ax_{n})x_{n} \\
        x_{n+1} & =ax_{n}-a(x_{n}x_{n})
\end{eqnarray}
\end{subequations}
all of which are, of course, equivalent mathematically. Here we
restrict ourselves to the expressions each of which can be written
in one program sentence, in other words, without using an extra
variable representing any term in the expressions above.

We are not the first to study expression dependence as well as machine
dependence of the computational results of chaotic systems. Colonna 
calculated iterations up to 80 steps for the forward-discretized (which
means using the Euler method of discretization) logistic differential equation, 
which is well known to be essentially equivalent (conjugate) to the logistic map,
on two different IBM machines using five different expressions \cite{eqdepend}.
He reported that 10 results are incompatible to one another even at the 80th iteration
for the case that corresponds to our results for  $a=4.0$.
Our studies here however, is more
specific, namely the effects of the computational modes $[53:53]$ and 
$[53:64]$, and more thorough, namely six different expressions
(L1)$\sim$(L6) over millions of iterations if necessary.

Oteo and Ros also compared the expressions essentially equivalent to our (L1)
and (L5) in their paper concerning errors when the logistic map is calculated
with double precision(DP) \cite{errors}. For $a=4.0$ they reported that the difference in the
mathematical expressions brings about the exponential growth of the distance between 
the two orbits even for the same initial conditions, namely after 50-60 iterations the
difference becomes as large as the size of the system itself.
Our concern here is again more thorough; we consider possible six different 
mathematical expressions as well as two different
computational modes in DP calculation not only for $a=4.0$
but also for other values of $a$.

For some special values of $a$, especially for $a=4.0$, six expressions
above are grouped into only two. As is well known for $a=4.0$ the
logistic map generates so-called fully developed chaos, hence very
many numerical calculations are done for this parameter value
to show typical results of the chaotic map. The value 4.0 can be realized
as an error-free number on any FPU's, hence multiplication by 4.0 does not
produce further errors in mantissa of the result. As the proof is given
in the Appendix, the six expressions of the logistic map are grouped as\\
\\
\begin{tabular}{ll}
\hspace{1.5cm}Group A: & \hspace{1cm}(L1), (L2), (L3)\\
\hspace{1.5cm}Group B: & \hspace{1cm}(L4), (L5), (L6)\\
\end{tabular}\\
\\
for $a=4.0$. 

The point we here make is that the value $a=4.0$ is quite particular as
far as numerical results are concerned, since for the values $2.0<a<4.0$
this grouping is no longer valid. Thus one must consider six different
expressions separately for the region. We show that six different results emerge for
six expressions even for the value of $a$ that is smaller than for
$4.0$ by only 1 bit in the next section.

\section{Virtual Periodicity} \label{mytt:VP}
In chaotic regime all the periodic orbits become unstable, hence they
should not be observed experimentally or numerically. However, when
computed on computers, chaotic orbits inevitably fall into virtual
periodic orbits due to restricted finite precision. For the logistic map
the virtual periodicity due to double precision(DP) has already been investigated \cite{perilen1}\cite{perilen2}\cite{perilen3}.
Keller and Wiese reported virtual periods for $a=3.90$ and $a=3.99$,
and Wang \textit {et al.} found the period length 5638349 of the most
frequently observed period for $a=4.0$. Wagner confirmed the same
period length as that of Wang \textit {et al.},  when he used Weitek, Mips and Sparc machines
but found a new length 86058417 as the most frequently observed when
calculation was done on VAX machines.  All the authors of
\cite{perilen1}\cite{perilen2}\cite{perilen3} however did not mention the computational modes or the
map expressions in their programs.

Results of our own calculations of virtual periods for $a=4.0$ are tabulated
in Table \ref{mytt:period}.
For $a=4.0$ only two expressions are needed to be compared as pointed out
\begin{table}[h]
\tbl{The virtual periods and their frequencies for two computational modes and
two expressions (a = 4.0). \label{mytt:period} } 
{\begin{tabular}{r r r r r r r r}\\[-2pt]
\toprule
\multicolumn{2}{c}{Group A $[$53:53$]$} & \multicolumn{2}{c}{Group A $[$53:64$]$} & \multicolumn{2}{c}{Group B $[$53:53$]$} & \multicolumn{2}{c}{Group B $[$53:64$]$}\\
\hline\\[-2pt]
 period  &  frequency & period  &  frequency & period  &  frequency & period  &  frequency\\
\hline\\[-2pt]
 5638349  &  678 &  (zero)  & 588 & 9458152  &  296 & 33525897  &  745\\
 (zero)  &  173 & 15784521    &  409 & 21739953  &  273 &  3210244  &  144\\
 14632801  &  89 &    1122211 &    3 & 17503666  &  217 & 17354121  &   56\\
  2441806 &   25 &           &  & (zero)  &  166 &  (zero)  &    52\\
 2625633 &    20 &           &  & 2857100  &   47 &  1176817  &    3\\
\botrule
\multicolumn{8}{l}{ Period (zero) indicates that the value of the iterates falls into 0. }\\
\multicolumn{8}{l}{ The results of the frequencies vary slightly depending  on the initial conditions.}\\
\end{tabular} 
}
\end{table}
at the end of the previous section, we used the expression (L1) as the
representative of group A and (L5) for group B.  One thousand initial
conditions $(x_{0})$ that evenly distributed in $0.000999\leq x_{0} \leq
0.999000$ which simulates the unit interval. Table \ref{mytt:period} lists the top
5 virtual-orbit frequencies when 1000 orbits are calculated for possible 4 computational
conditions. From our results we can identify the
period 5638349 that was reported in \cite{perilen2} and \cite{perilen3} had been obtained when
computational mode $[53:53]$ and the map expression of group A were used.
From Table \ref{mytt:period} it is obvious that the virtual period length and its
frequency are strongly dependent on the computational mode and 
the map expression, therefore, if one wants to discuss virtual periods
due to DP, these conditions should be clarified for reproducibility of
the results.

If the value of $a$ is different from 4.0, six expressions of the logistic
map can no longer be grouped into two. We have hence 12 different virtual
periods as tabulated in Table \ref{mytt:periodm1b} for the value of the parameter 
$a$ whose mantissa is
only 1 bit smaller than 4.0, i.e., 3.9999999999999995 in decimal or
0x400fffffffffffff in hexadecimal. This value is the largest possible value
smaller than 4.0 that can be expressed in DP. One thousand orbits were
also surveyed and the most frequent virtual period is tabulated in
Table \ref{mytt:periodm1b} for each category. Notice that no two entries become identical
in Table \ref{mytt:periodm1b}, which means that once the parameter value of $a$ 
is smaller than 4.0 even by just one bit, the virtual periodic phenomena become
quite complicated. Furthermore, no entries in Table \ref{mytt:periodm1b} become identical
to any of those in Table \ref{mytt:period}.\\

\begin{table}[h]
\tbl{The virtual periods and their frequencies for two computational modes and
six expressions (a = 0x400fffffffffffff in hexadecimal). \label{mytt:periodm1b}}
{\begin{tabular}{c r r r r r}\\[-2pt]
\toprule
\multicolumn{1}{c}{} & \multicolumn{2}{c}{[53:53]} & \multicolumn{2}{c}{[53:64]} \\
expression & period  &  frequency & period  &  frequency  \\
\colrule
 L1 & 144666122  &  973 & 27919860 & 943\\
 L2 & 133933248  &  922 & 34082242  & 999  \\
 L3 & 22875200 & 805 &  14808396 &  900\\
 L4 & 9726075  &  894 & 46556242  &  776 \\
 L5 & 9440450  & 845  & 62494847  & 719 \\
 L6 & 26643051  &  966  & 25600176  &  879 \\
\botrule
\end{tabular}}
\end{table}
Although emergence of the virtual periods is quite complicated, they are very 
long---even the shortest is of the order of $10^{6}$---and are seldom observed
in usual computer experiments. But we found that in more familiar results such as
the bifurcation diagram the effects of the six different expressions due to the
difference in the computational modes become apparent for $a<4.0$, as will be presented
in the next two sections.

\section {Difference in Bifurcation Diagrams}	\label{mytt:DBD}
For a continuous range of the external parameter $a$, the
bifurcation diagrams are most commonly drawn to grasp dynamical
behaviors of the system. In this section, in order to show that the bifurcation diagram
is actually dependent on the expressions (L1)$\sim$(L6) and the computational
modes of  FPU, namely, $[53:53]$ or $[53:64]$, we pay attention to the
difference between two bifurcation diagrams calculated with different expressions and
computational modes. Since there are as many as $6!/(2!4!) =15$ ways
to select two expressions from the expressions (L1)$\sim$(L6), and each 
pair must be computed with the computational mode $[53:53]$ and $[53:64]$,
we first use the expression (L1) and (L5) as representatives for our
exposition below. These two are the most frequently employed expressions
in literatures because when one expresses Eq.(\ref{mytt:eq1}) directly in program
language it becomes (L1), and (L5) is obtained when the expanded form of
Eq.(\ref{mytt:eq1}), i.e, $ax-ax^{2}$ is directly programed. After the explanation,
though, results for all the 30 combinations are presented.

The bifurcation diagram is drawn as follows: a range of interest for
the external parameter $a$, specifically $a_{min} \leq a \leq a_{max}$
is divided evenly by a positive integer $M$, for each value of $a$
the iterated values of the map starting from the initial value $x_{0}$
are plotted. In most studies behaviors of the system after sufficiently
many iterations attract attentions, hence the values of the first
$I_{d}$ iterations are not plotted and ensuing $(I_{t}-I_{d})$ iterations
are plotted. Here $I_{t}$ is the total number of iterations for each
value of $a$. Although seldom shown in literatures, it is possible to
observe transient behaviors of the map if one plots all the iterations
by putting $I_{d}=0$.

Now, let us begin the exposition of our investigation using (L1) and (L5).
It is quite natural to guess that bifurcation diagrams are the same for all the
mathematically equivalent expressions (L1)$\sim$(L6), as long as the
parameters $a_{mim}$, $a_{max}$, $M$, $x_{0}$, $I_{d}$ and $I_{t}$ are
all unaltered. However, we show that the difference emerges depending on
the expression and the mode of computation. Furthermore, we show that
it is possible to evaluate the difference quantitatively.

\begin{figure}[h]
  \begin{tabular}{lll}
    \begin{minipage}[t]{0.3\textwidth}
      \includegraphics[width=\textwidth]{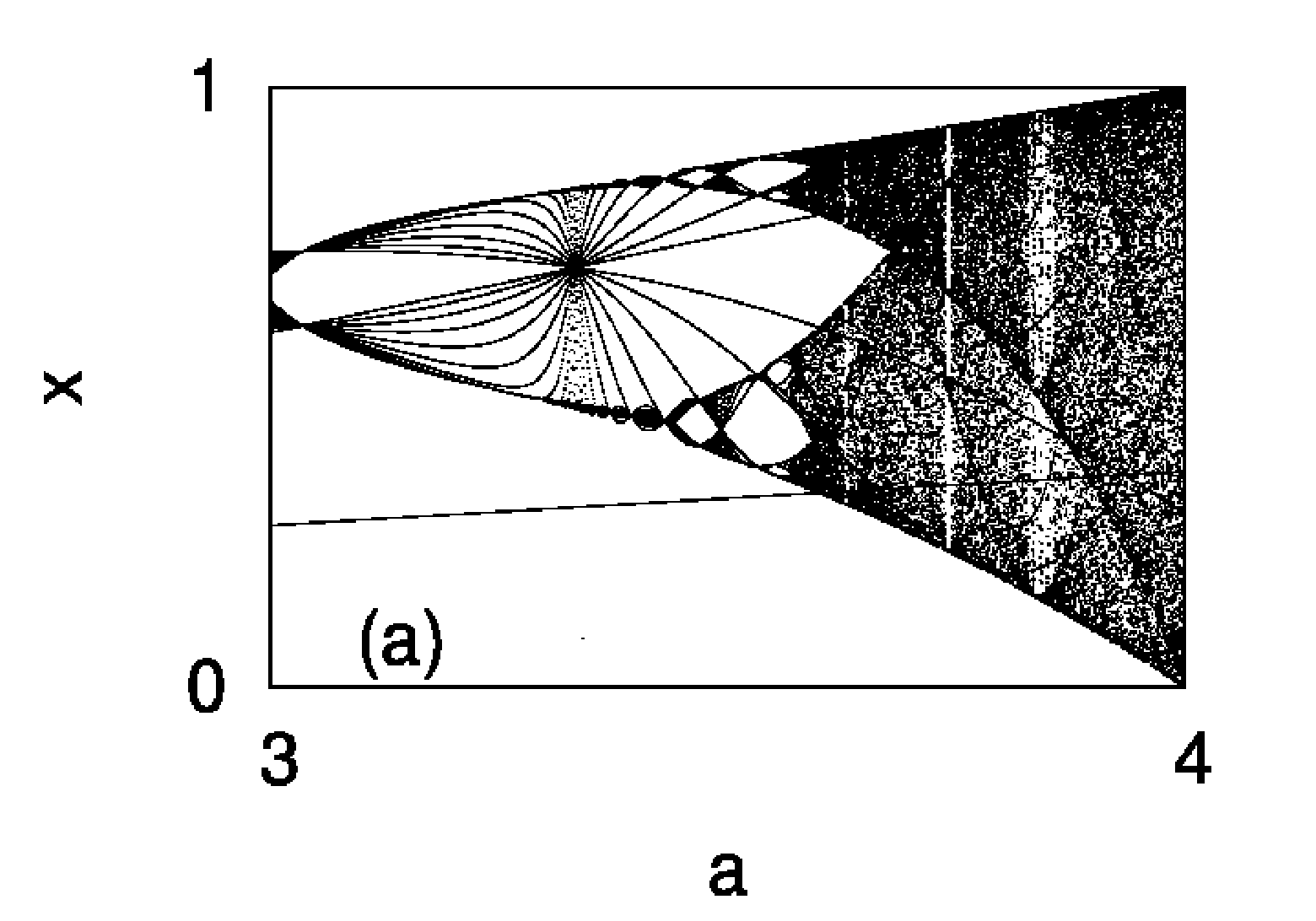}
    \end{minipage} &

    \begin{minipage}[t]{0.3\textwidth}
      \includegraphics[width=\textwidth]{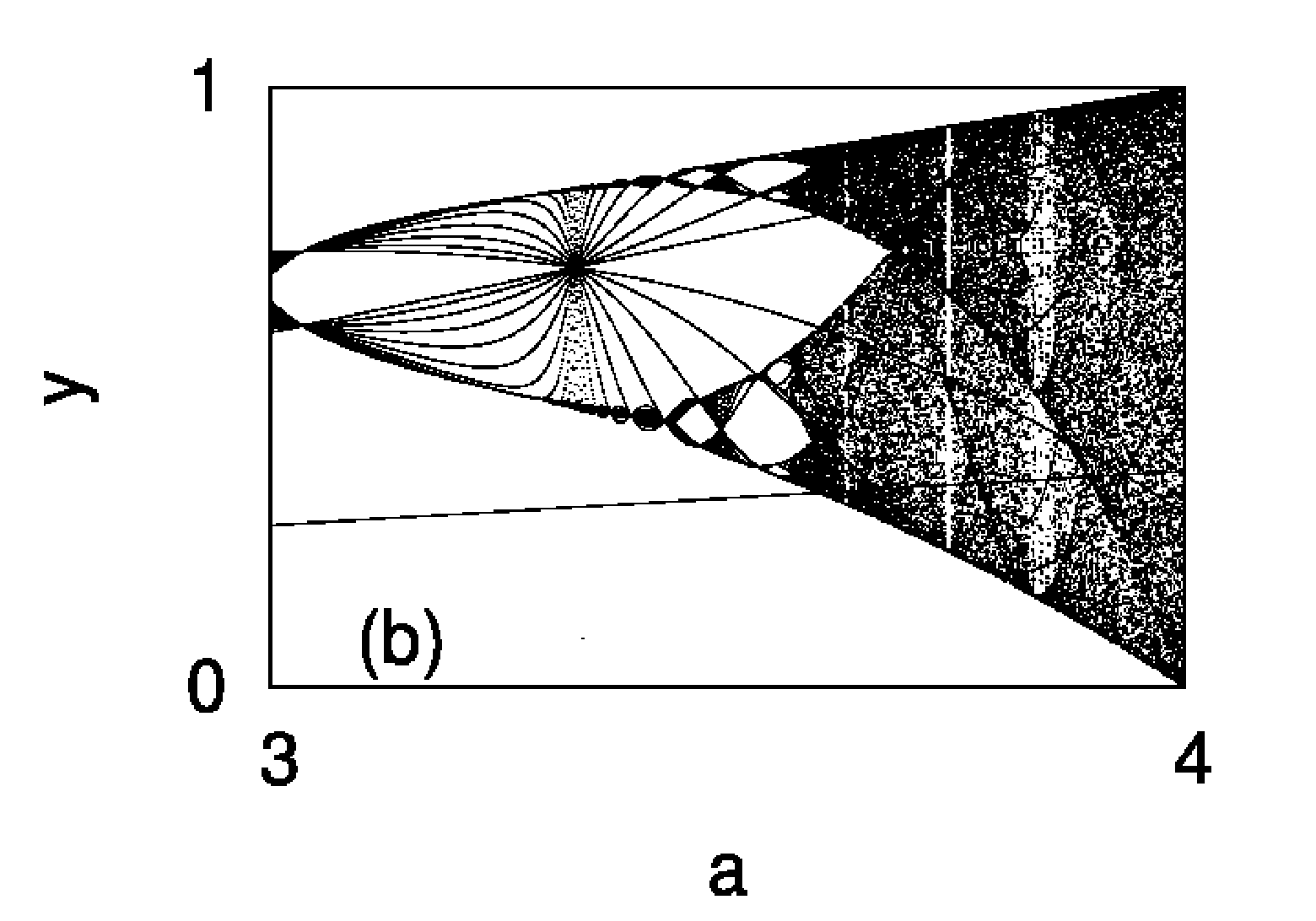}
    \end{minipage} &

    \begin{minipage}[t]{0.3\textwidth}
      \includegraphics[width=\textwidth]{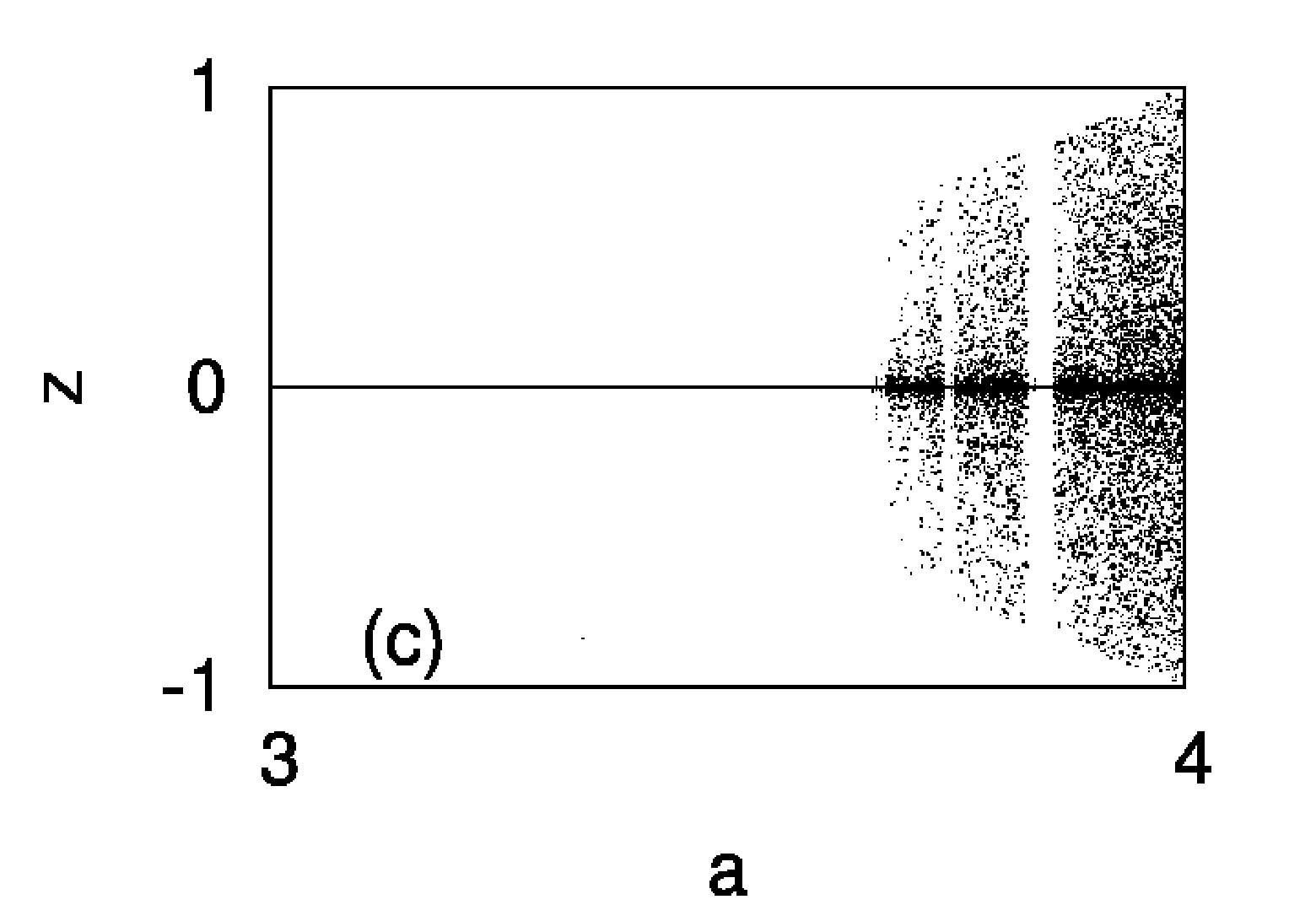}
    \end{minipage}
  \end{tabular}
\caption{The bifurcation diagrams for the computational mode $[53:53]$ with the
parameter values $a_{min}=3.0$, $a_{max}=4.0$, $M=1000$,
$x_{0}=0.1$, $I_{d}=0$ and $I_{t}=100$. (a) The bifurcation diagram for the expression
(L1). (b) The bifurcation diagram for (L5). (c) The bifurcation diagram for the difference Eq.(\ref{mytt:eq7})
where $x_{n}$ stands for (L1) and $y_{n}$ for (L5).}
      \label{mytt:fig1}
\end{figure}

\begin{figure}[h]
  \begin{tabular}{lll}
    \begin{minipage}[t]{0.3\textwidth}
      \includegraphics[width=\textwidth]{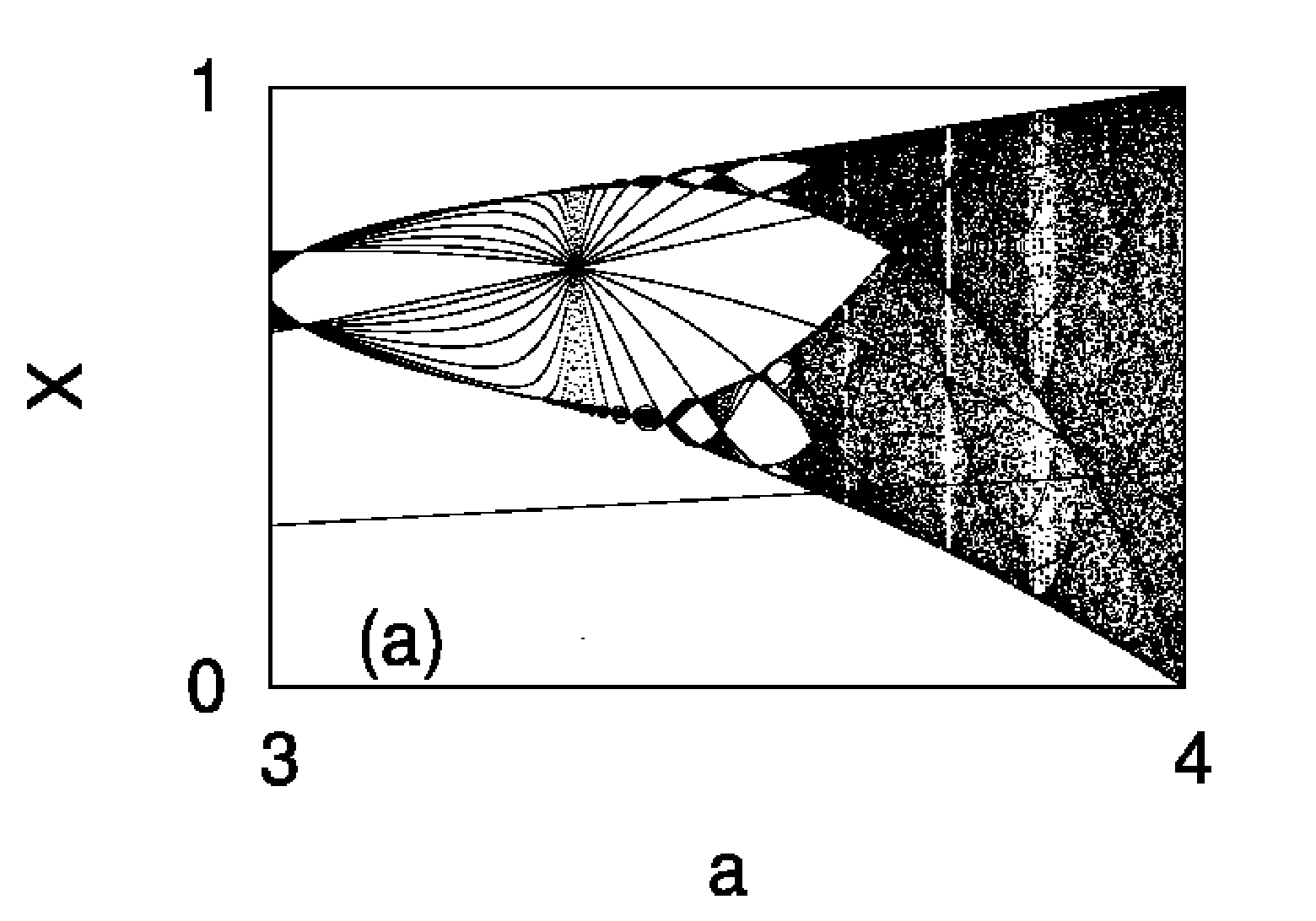}
    \end{minipage} &

    \begin{minipage}[t]{0.3\textwidth}
      \includegraphics[width=\textwidth]{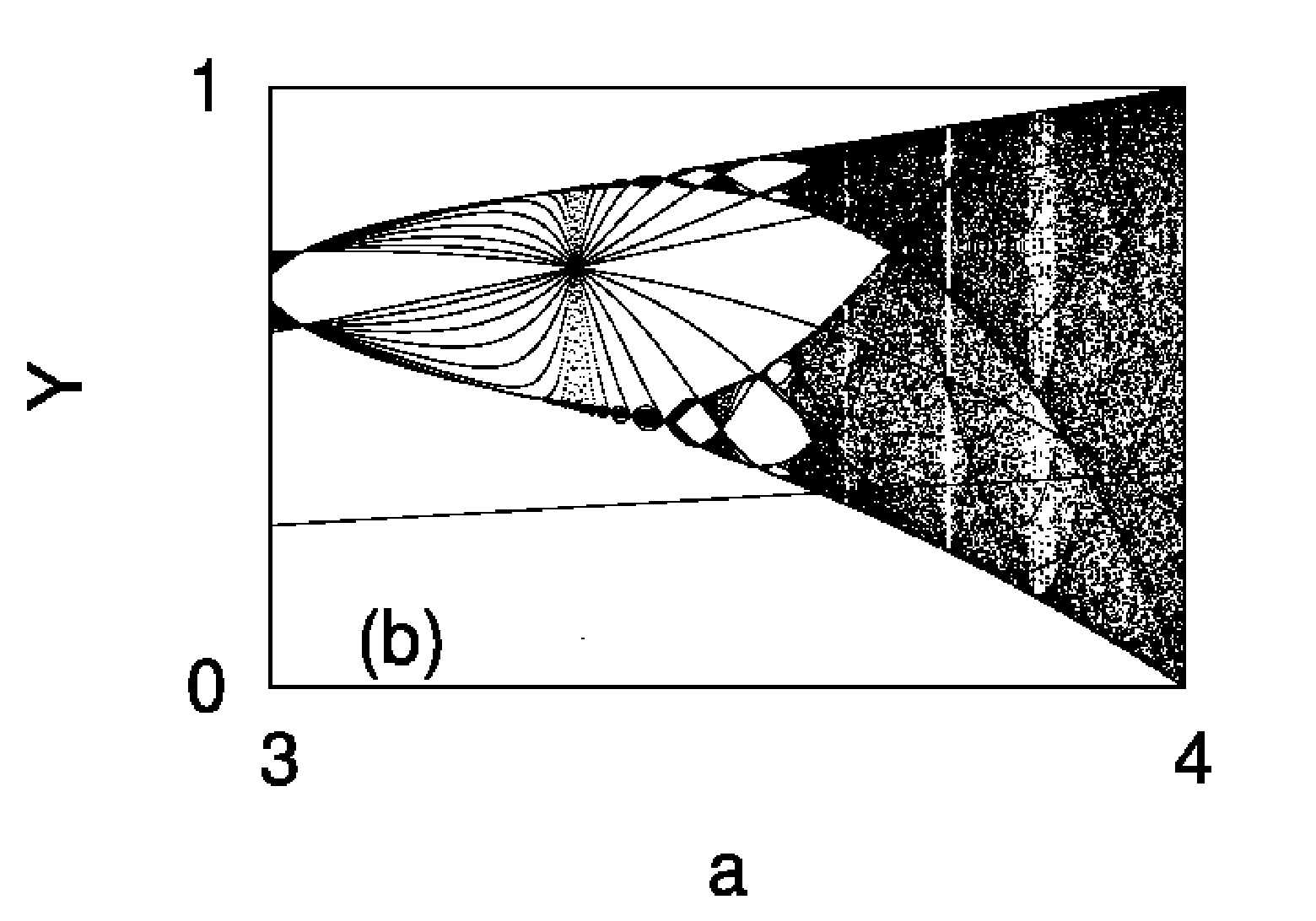}
    \end{minipage} &

    \begin{minipage}[t]{0.3\textwidth}
      \includegraphics[width=\textwidth]{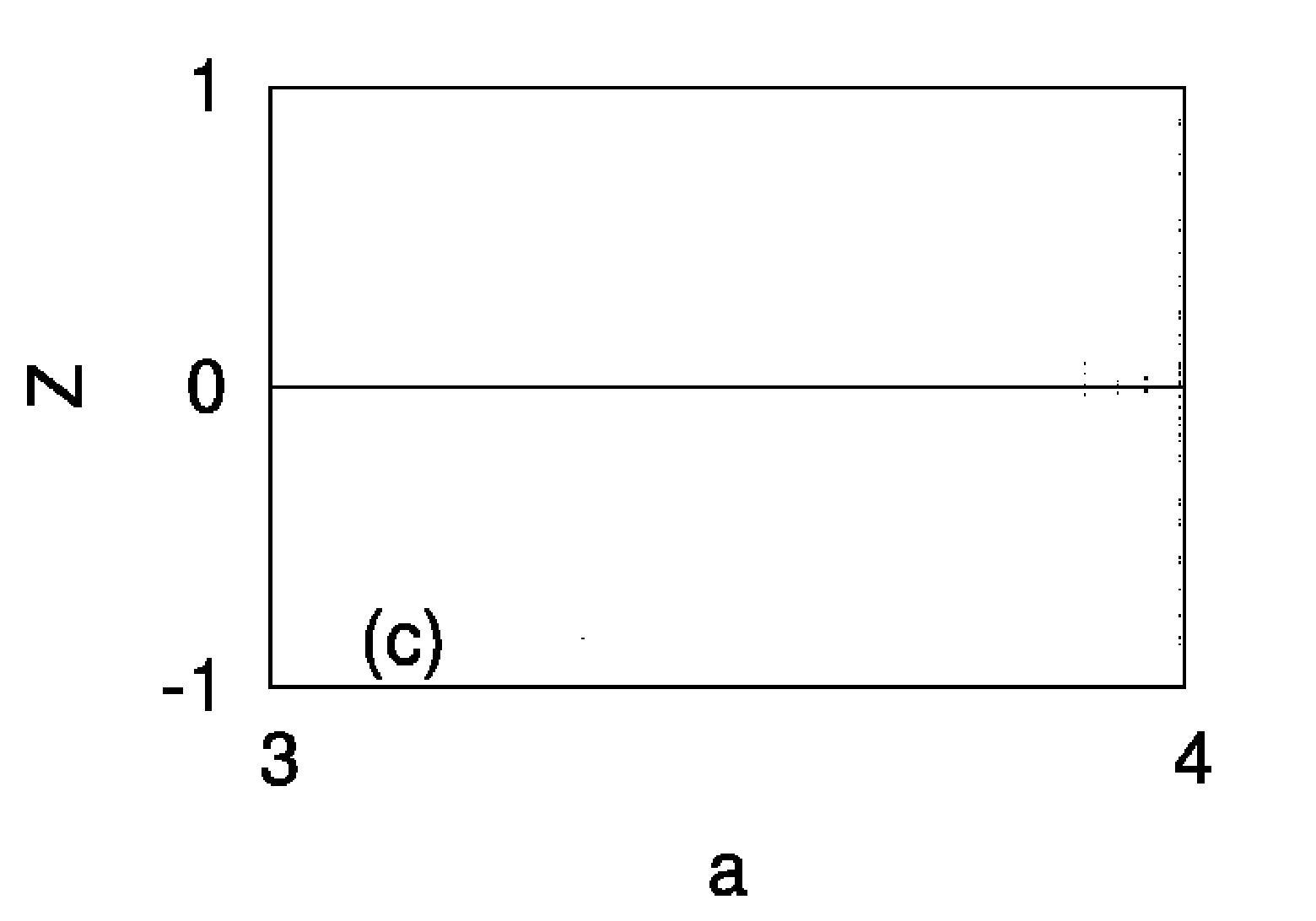}
    \end{minipage}

  \end{tabular}
\caption{The bifurcation diagrams for the computational mode $[53:64]$ with the same
parameter values as those of Fig.\ref{mytt:fig1}. (a) The bifurcation diagram for the expression (L1).
(b) The bifurcation diagram for (L5). (c) The bifurcation diagram for the difference Eq.(\ref{mytt:eq8})
where $X_{n}$ stands for (L1) and $Y_{n}$ for (L5).}
      \label{mytt:fig2}
\end{figure}

To make our discussion transparent we use the variable $x$ when we run
the expression (L1) with $[53:53]$ mode, namely,
\begin{equation}
	x_{n+1}=(ax_{n})(1.0-x_{n}),		\label{mytt:eq3}
\end{equation}
and we use $y$ for the expression (L5) with $[53:53]$ mode as
\begin{equation}
	y_{n+1}=ay_{n}-(ay_{n})y_{n}.	 \label{mytt:eq4}
\end{equation}
We employ the upper-case letters $X$ and $Y$ for (L1) with $[53:64]$ mode as
\begin{equation}
	X_{n+1}=(aX_{n})(1.0-X_{n}),		\label{mytt:eq5}
\end{equation}
and (L5) with $[53:64]$ as
\begin{equation}
	Y_{n+1}=aY_{n}-(aY_{n})Y_{n}.	\label{mytt:eq6}
\end{equation}
Our first results are presented in Figs.\ref{mytt:fig1} and \ref{mytt:fig2}, where the parameter
values are fixed as $a_{mim}=3.0$, $a_{max}=4.0$, $M=1000$, $x_{0}=0.1$,
$I_{d}=0$ and $I_{t}=100$, in other words, the first 100 iterations
starting from $x_{0}=0.1$ are plotted for 1000 evenly separated values
of $a$ from 3.0 to 4.0. The bifurcation diagram for Eq.(\ref{mytt:eq3}) is shown in Fig.\ref{mytt:fig1}(a), Eq.(\ref{mytt:eq4}) in
Fig.\ref{mytt:fig1}(b), Eq.(\ref{mytt:eq5}) in Fig.\ref{mytt:fig2}(a) and Eq.(\ref{mytt:eq6}) in Fig.\ref{mytt:fig2}(b). To our naked eyes
all the four diagrams appear to be similar, but comparing  Fig.\ref{mytt:fig1}(c)
with Fig.\ref{mytt:fig2}(c) one immediately sees that there really is a difference.
In Fig.\ref{mytt:fig1}(c) the difference
\begin{equation}
	z_{n}=x_{n}-y_{n}			\label{mytt:eq7}
\end{equation}
and in Fig.\ref{mytt:fig2}(c) the difference
\begin{equation}
	Z_{n}=X_{n}-Y_{n}			\label{mytt:eq8}
\end{equation}
is plotted, respectively. The bifurcation diagram in Fig.\ref{mytt:fig1}(a) is different from that
in Fig.\ref{mytt:fig1}(b) in chaotic regime, whereas the bifurcation diagrams in Fig.\ref{mytt:fig2}(a) and Fig.\ref{mytt:fig2}(b)
are the same at least up to $I_{t}=100$.

It is understandable that there are no differences in the period-doubling
region $(3.0 \leq a \lesssim 3.57)$, but remarkable results are for the
chaotic region $(3.57 \lesssim a \leq 4.0)$, namely  $z_{n}$ spreads
chaotically (Fig.\ref{mytt:fig1}(c)) whereas $Z_{n}$ remains 0 (Fig.\ref{mytt:fig2}(c)).\footnote{If
one carefully looks at Fig.\ref{mytt:fig2}(c) some scattered points near $a \lesssim 4.0$
are to be recognized.} Chaotic behavior of $z_{n}$ for this region can
be understood that for the mode $[53:53]$ the orbit calculated with
(L1) becomes different from that with (L5) within 100 iterations, however,
the fact that $Z_{n}$ stays at 0 reflects the fact that for the
mode $[53:64]$ despite the fact that both $X_{n}$ and $Y_{n}$ in Eq.(\ref{mytt:eq8})
dance chaotically the values of them coincide completely within the double
precision(DP) up to 100 iterations. Figs.\ref{mytt:fig1}(c) and \ref{mytt:fig2}(c) show
clearly that the computational mode does affect chaotic orbits.

\begin{figure}[h]
  \begin{tabular}{lll}
    \begin{minipage}[t]{0.3\textwidth}
      \includegraphics[width=\textwidth]{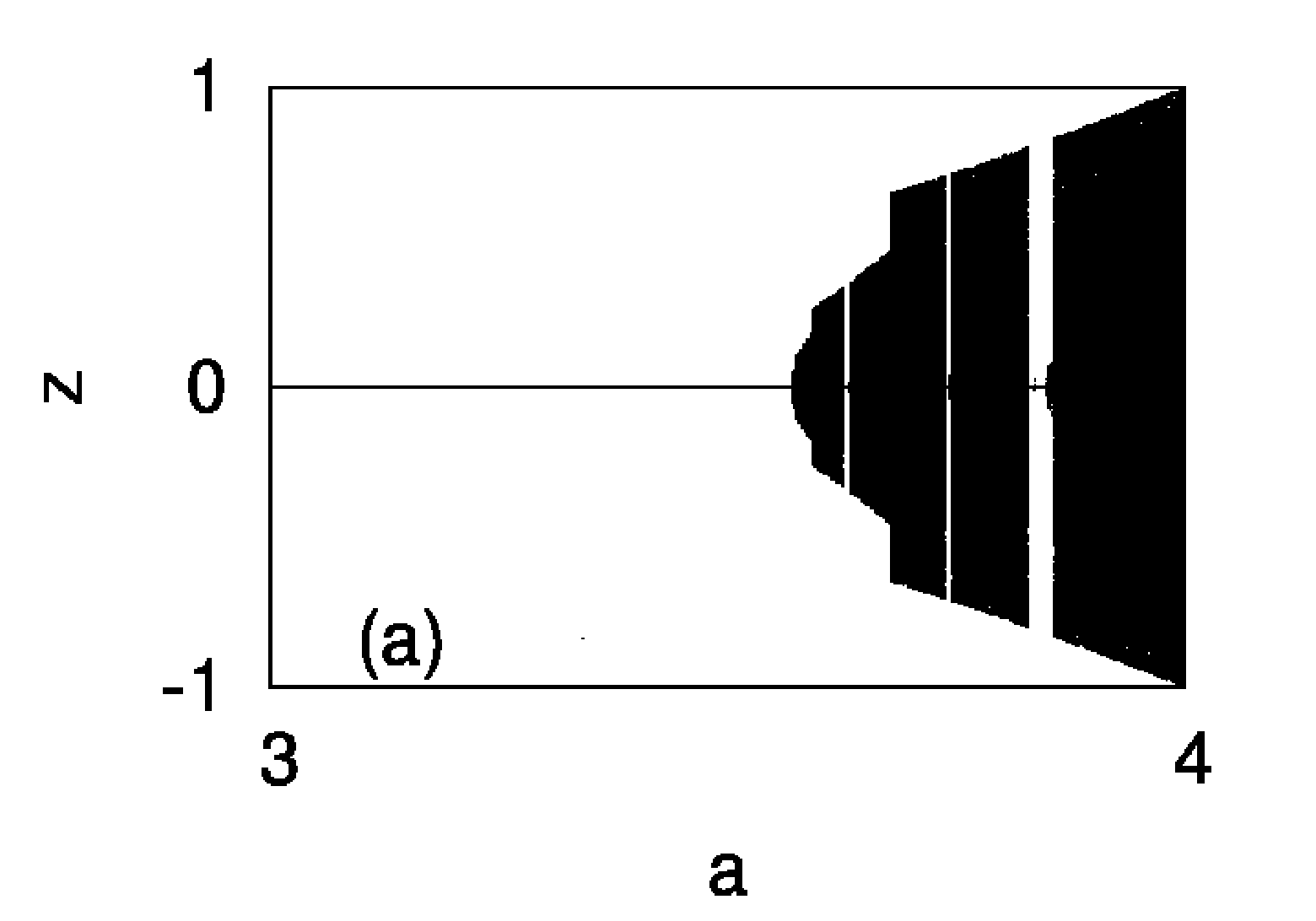}
    \end{minipage} &

    \begin{minipage}[t]{0.3\textwidth}
      \includegraphics[width=\textwidth]{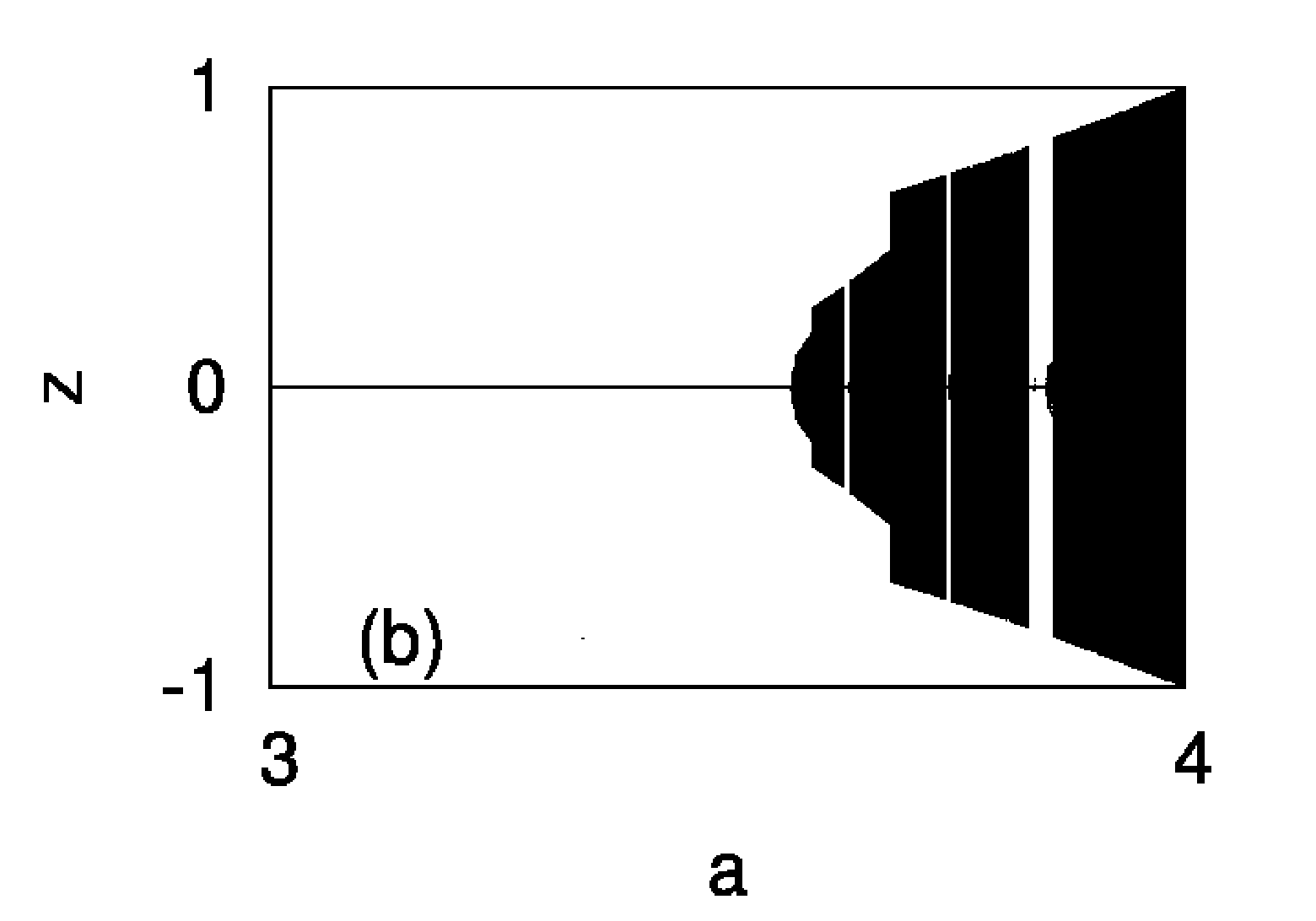}
    \end{minipage} &

    \begin{minipage}[t]{0.3\textwidth}
      \includegraphics[width=\textwidth]{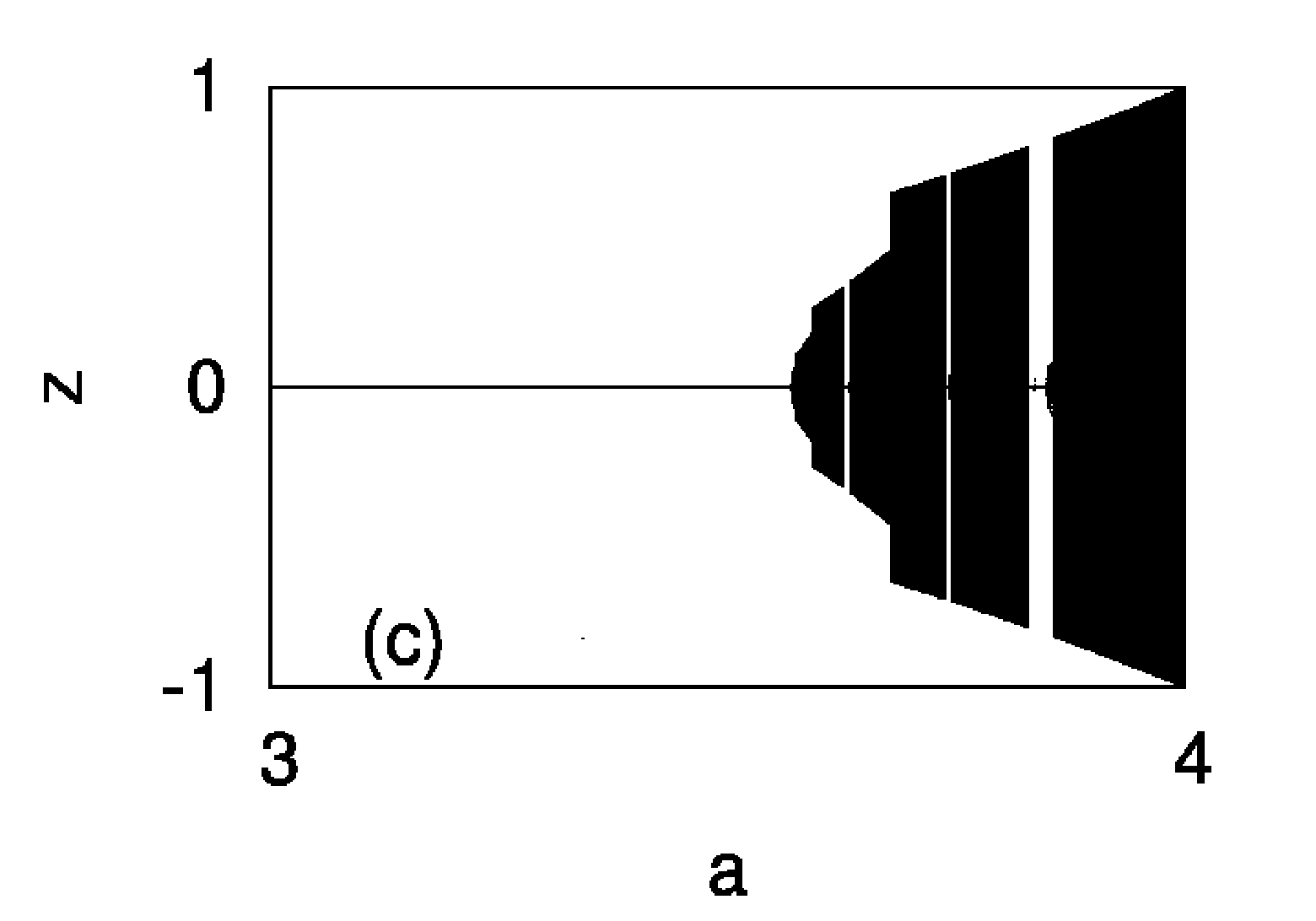}
    \end{minipage}

  \end{tabular}
\caption{The bifurcation diagrams for the computational mode $[53:53]$ for the
difference Eq.(\ref{mytt:eq7}) for expressions (L1) and (L5). (a) $I_{t}=1000$.
(b) $I_{t}=3000$. (c) $I_{t}=10000$.  Other parameter values are the
same as those used for Fig.\ref{mytt:fig1}.}
      \label{mytt:fig3}
\end{figure}

\begin{figure}[h]
  \begin{tabular}{lll}
    \begin{minipage}[t]{0.3\textwidth}
      \includegraphics[width=\textwidth]{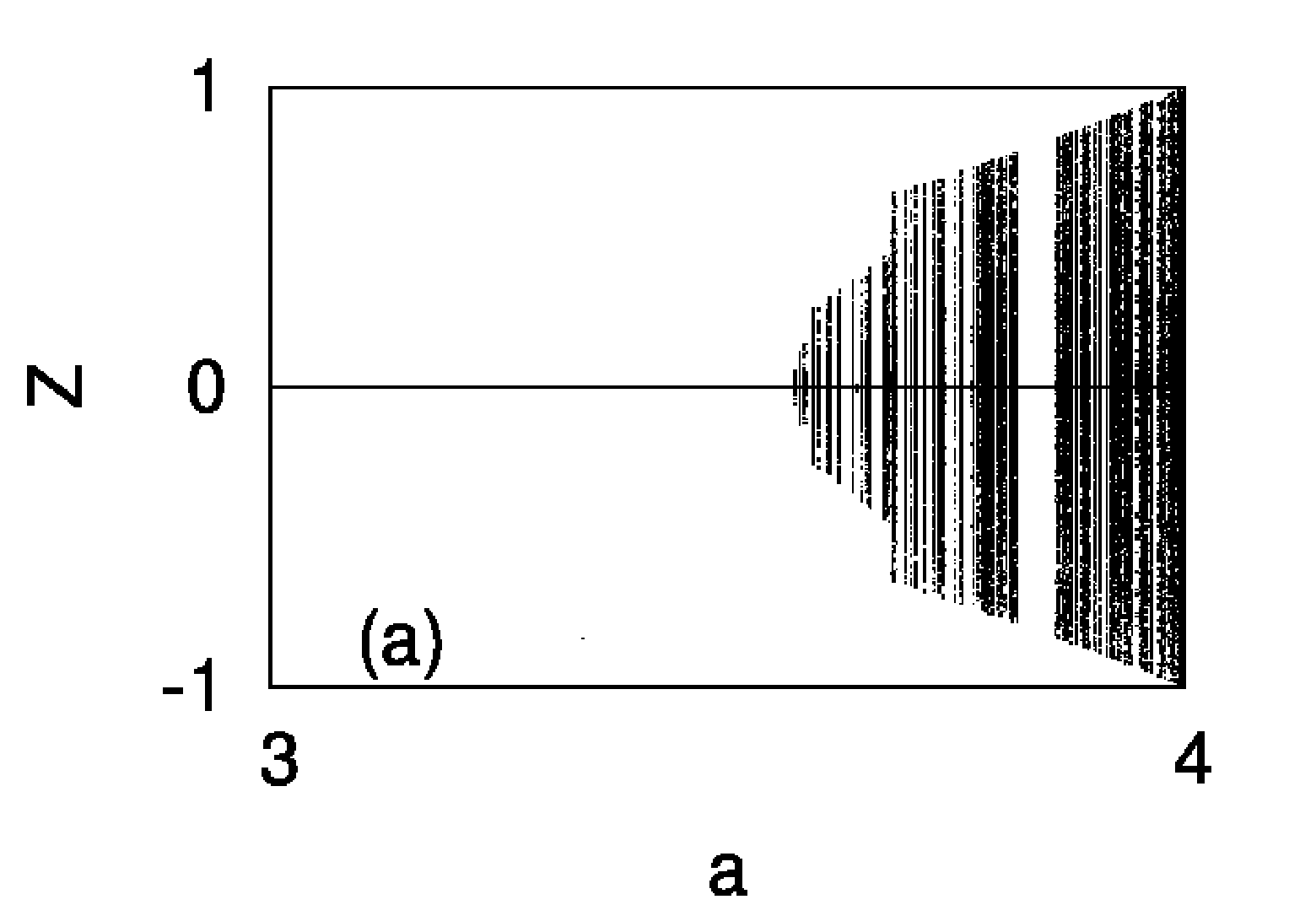}
    \end{minipage} &

    \begin{minipage}[t]{0.3\textwidth}
      \includegraphics[width=\textwidth]{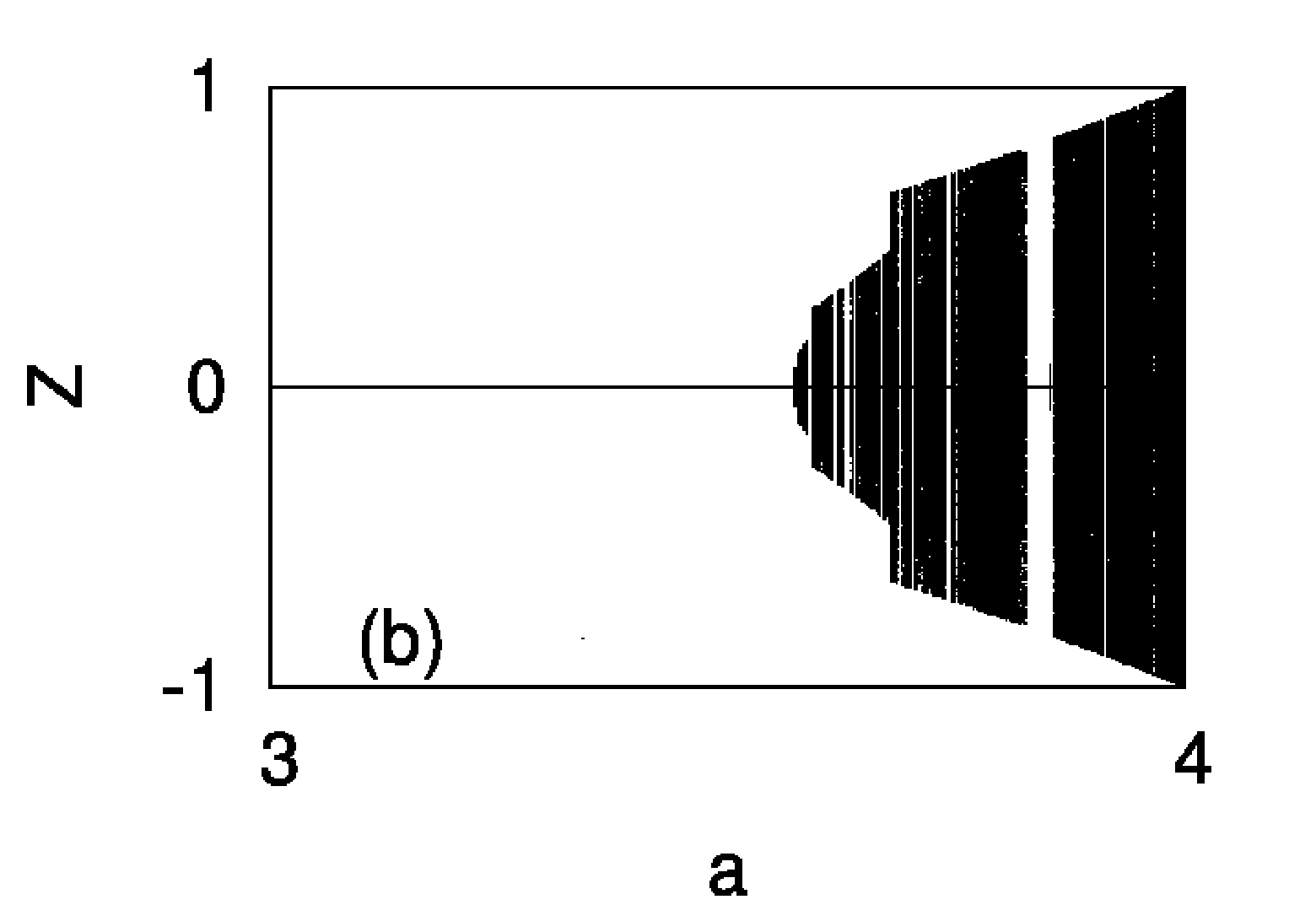}
    \end{minipage} &

    \begin{minipage}[t]{0.3\textwidth}
      \includegraphics[width=\textwidth]{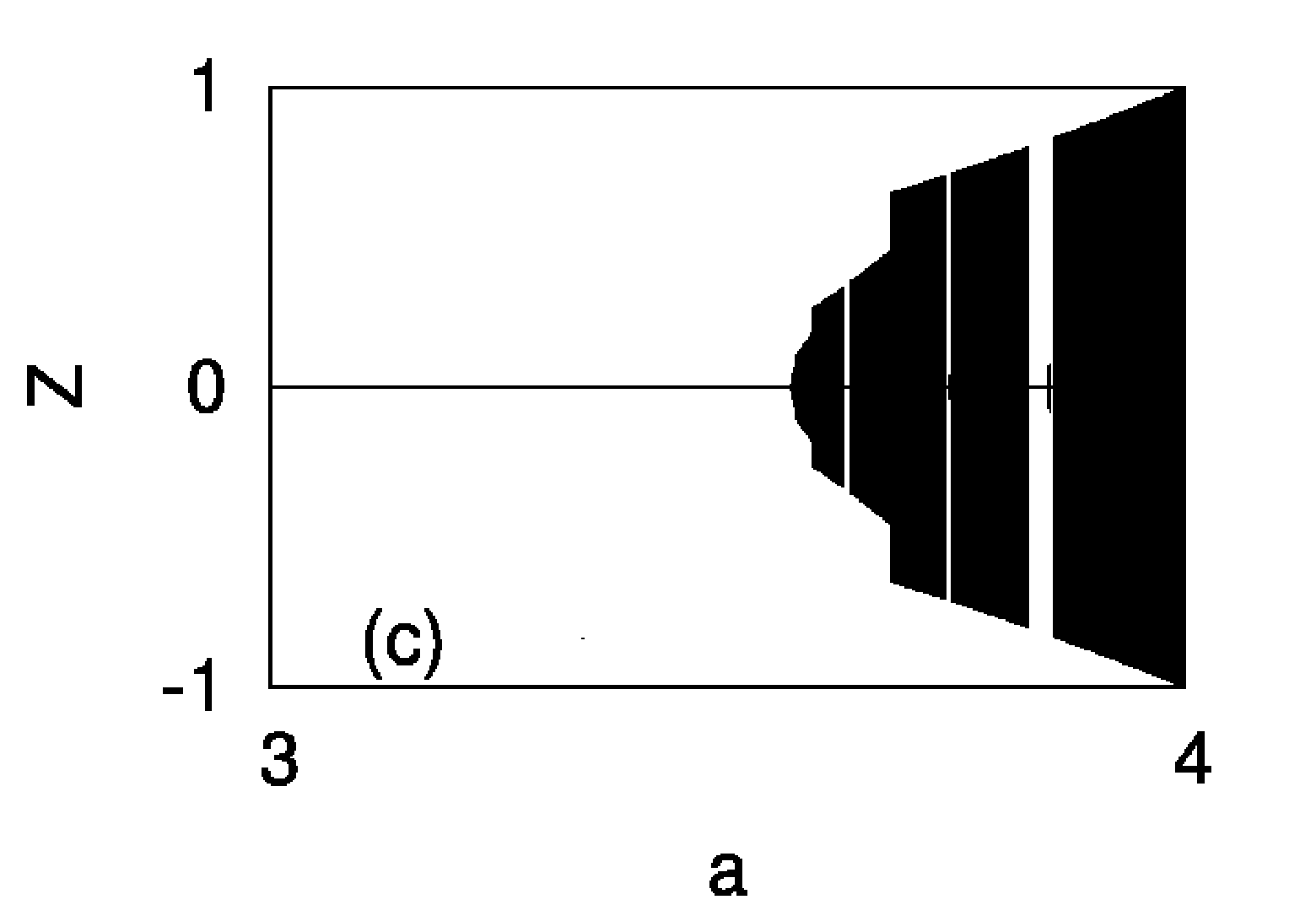}
    \end{minipage}

  \end{tabular}
\caption{The bifurcation diagrams for the computational mode $[53:64]$ for the
difference Eq.(\ref{mytt:eq8}) for expressions (L1) and (L5). (a) $I_{t}=1000$.
(b) $I_{t}=3000$. (c) $I_{t}=10000$.  Other parameter values are the
same as those used for Fig.\ref{mytt:fig1}.}
      \label{mytt:fig4}
\end{figure}

Up to this point the total number of iterations $I_{t}$ is restricted to
100. Now let us increase $I_{t}$ as $I_{t}=1000, 3000$ and 10000, and see
what happens to $z_{n}$ (Fig.\ref{mytt:fig3}) and to $Z_{n}$ (Fig.\ref{mytt:fig4}). If we compare Fig.\ref{mytt:fig3}(b)
with Fig.\ref{mytt:fig4}(b), we see that $z_{n}$ is still different from $Z_{n}$. However,
from Fig.\ref{mytt:fig3}(c) and Fig.\ref{mytt:fig4}(c) we cannot recognize any conspicuous differences.
The observational results stated above (independence of the initial value
$x_{0}$ has been confirmed), that for the mode $[53:53]$ the difference due
to expressions (L1) and (L5) disappears by $I_{t}=1000$,
but for the mode $[53:64]$ there exist certain values of $a$ for which
$X_{n}$ and $Y_{n}$ take the same value even after 1000 iterations.
For $I_{t}=10000$, however, the bifurcation diagram for $Z_{n}$ cannot be distinguished
from that for $z_{n}$ (Figs.\ref{mytt:fig3}(c) and \ref{mytt:fig4}(c)).

In the following we try to quantify the difference between the bifurcation diagrams
for $z_{n}$ and $Z_{n}$.  Since the difference becomes 0 for the
periodic region, we consider only the chaotic region putting
$a_{min}=3.5699456718$ which is the available value \cite{timeseries} for the accumulation
point of the period doubling or the Feigenbaum point and $a_{max}=4.0$.

\begin{figure}[h]
  \begin{tabular}{cc}
    \begin{minipage}[b]{0.5\textwidth}
      \includegraphics[width=\textwidth]{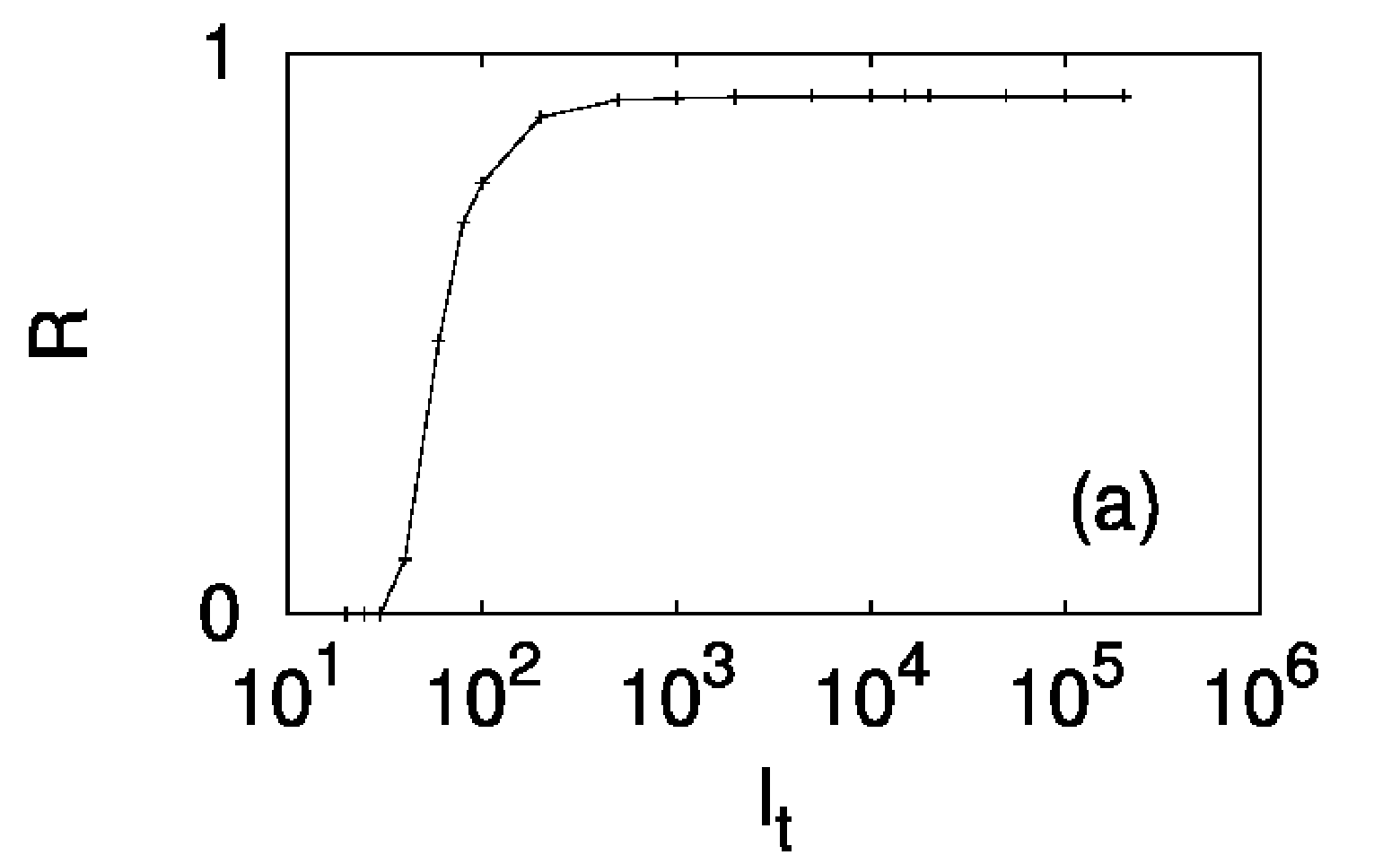}
    \end{minipage}

    \begin{minipage}[b]{0.5\textwidth}
      \includegraphics[width=\textwidth]{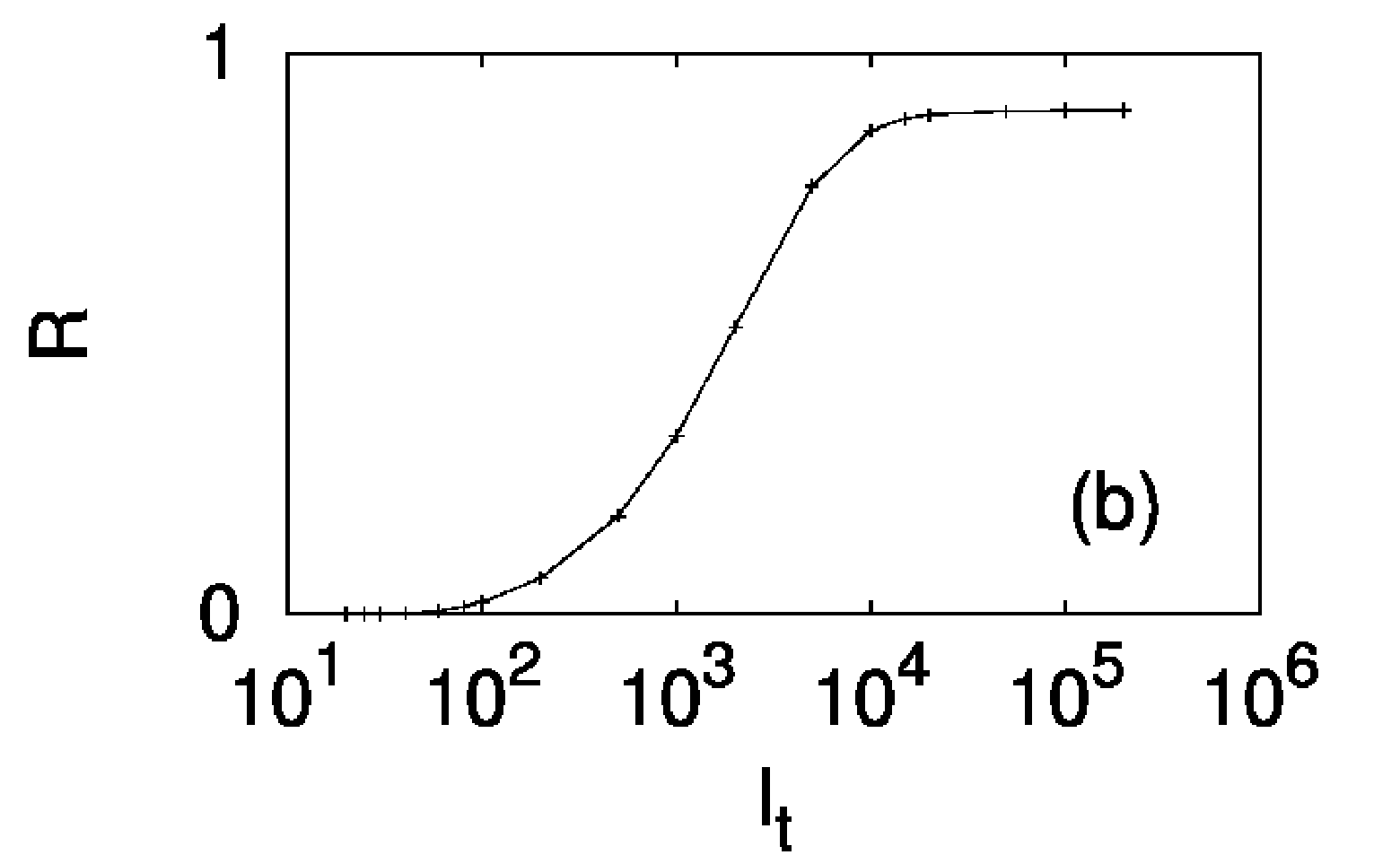}
    \end{minipage}
  \end{tabular}
\caption{The graph of the ratio of chaotic orbits $R(I_{t})$ defined by
Eq.(\ref{mytt:eq9}) against $I_{t}$ calculated for the difference between expressions
(L1) and (L5). (a) The mode $[53:53]$. (b) The mode $[53:64]$.}
      \label{mytt:fig5}
\end{figure}

First we divide $(a_{max}-a_{min})$ by $M$ and for each value of $a$
we prepare evenly distributed $P$ initial values of $x_{0}$. The total
iteration number is $I_{t}$. If the absolute value $|z_{n}|$ or
$|Z_{n}|$ exceeds a certain threshold value $\epsilon$, it can grow
exponentially afterwards, so the corresponding point $(a,x_{0})$ can
be judged to give a chaotic orbit. If the total number of  points thus
determined to produce chaotic orbits up until $I_{t}$ is denoted by
$S(I_{t})$, then
\begin{equation}
	R(I_{t})=\frac{S(I_{t})}{MP}	\label{mytt:eq9}
\end{equation}
represents the ratio of the chaotic orbits up until $I_{t}$ iterations.
In Fig.\ref{mytt:fig5}(a) the ratio $R(I_{t})$ is plotted against $I_{t}$ for the mode
$[53:53]$. The number of initial points is $P=1000$ so that the region
$0.000999 \leq x_{0} \leq 0.999000$ divided by $P$ generate 1000 orbits.
In Fig.\ref{mytt:fig5}(b), $R(I_{t})$ is plotted against $I_{t}$ for the $[53:64]$ mode.
The threshold value $\epsilon$ is set at $10^{-6}$ for both graphs. Note
that these two graphs are different: for the $[53:53]$ mode, a sharp
transition occurs for $I_{t}<10^{2}$, whereas for $[53:64]$
mode a far gentler transition takes place whose mid-point is seen for
$I_{t}>10^{3}$.

\begin{figure}[h]
  \begin{tabular}{cc}
    \begin{minipage}[b]{0.5\textwidth}
      \includegraphics[width=\textwidth]{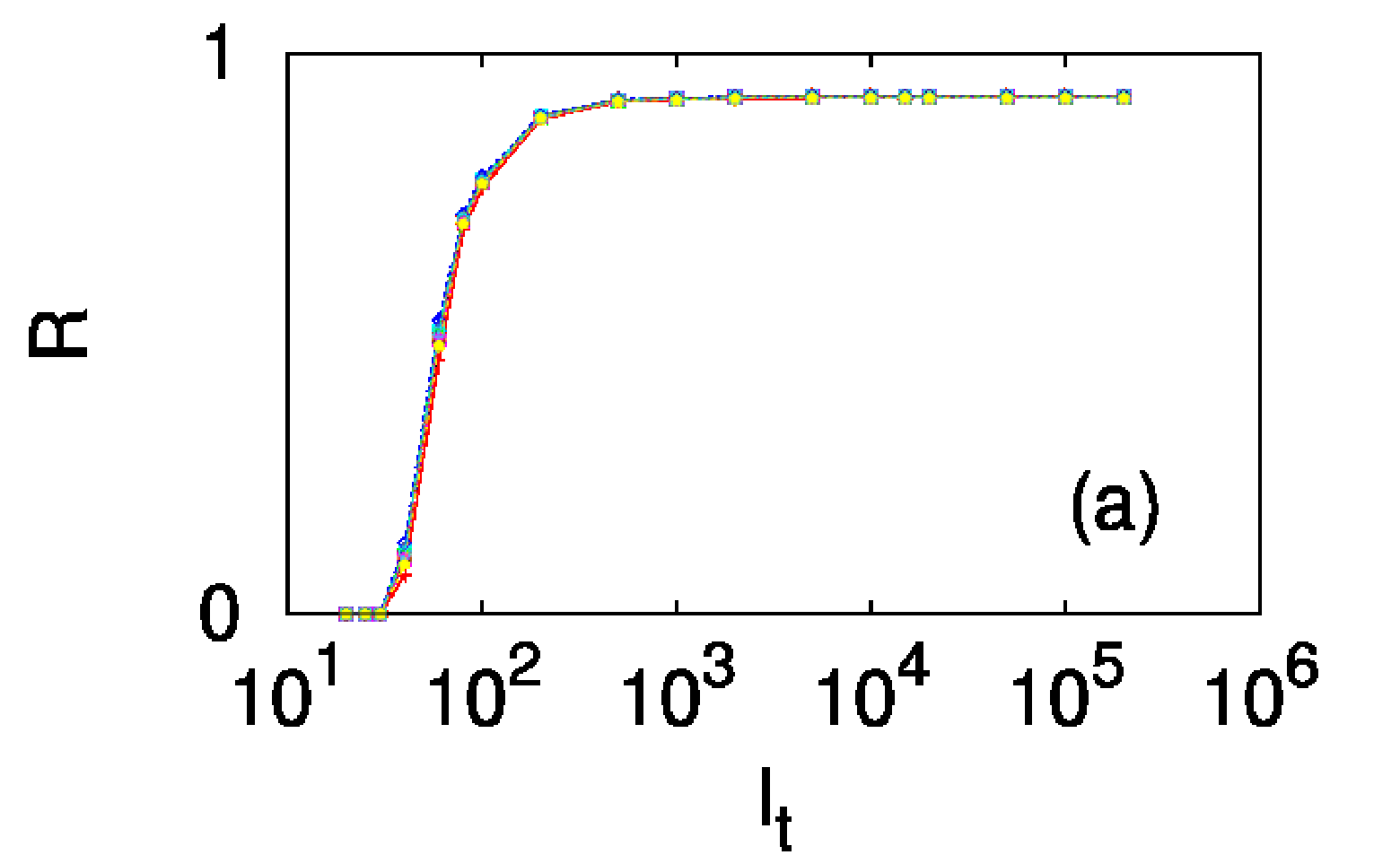}
    \end{minipage}

    \begin{minipage}[b]{0.5\textwidth}
      \includegraphics[width=\textwidth]{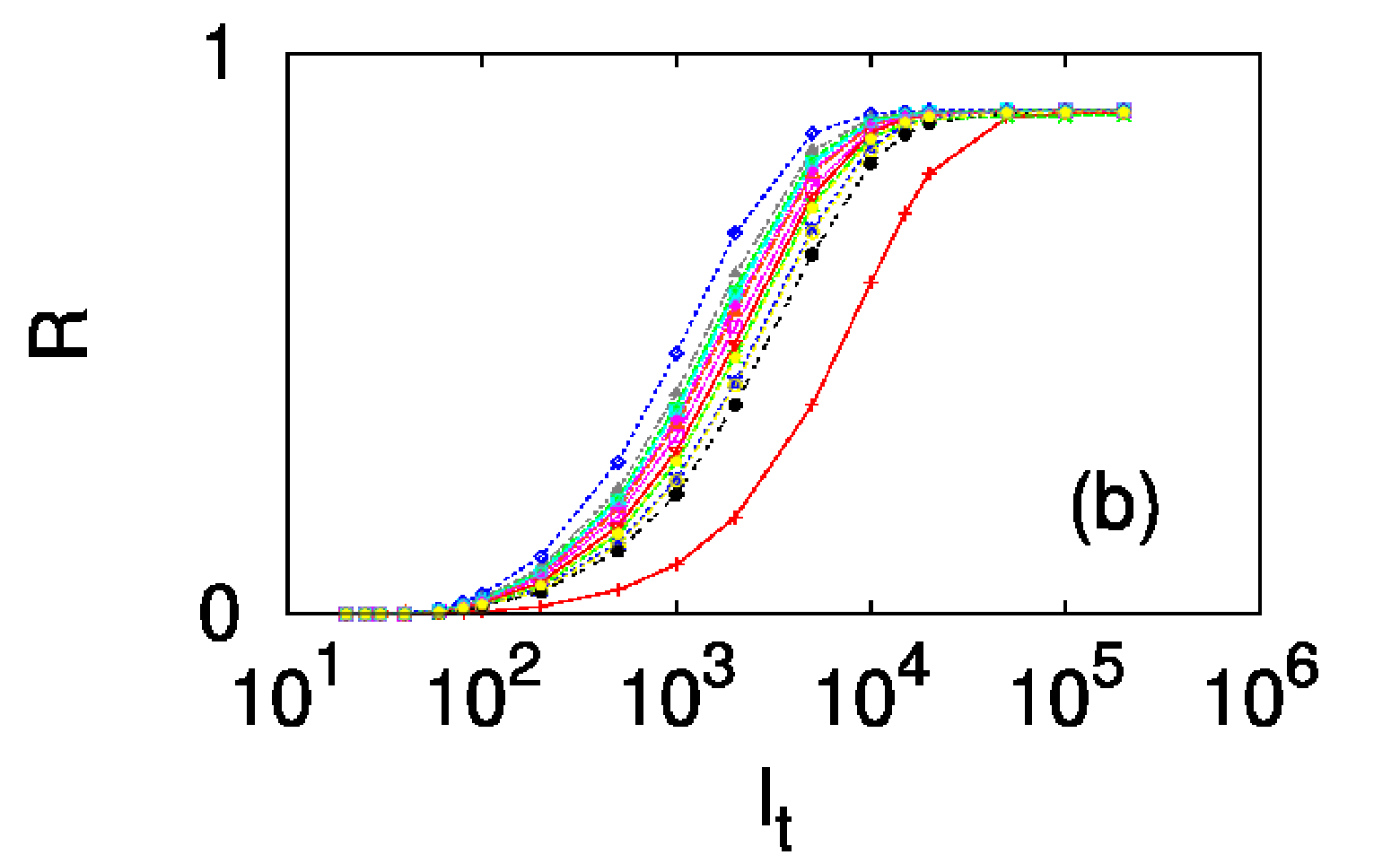}
    \end{minipage}
  \end{tabular}
\caption{The ratio of chaotic orbits $R(I_{t})$ for all the 15 pairs
selected from 6 different mathematical expressions against $I_{t}$.
(a) The mode $[53:53]$. (b) The mode $[53:64]$. Each curve respectively represents the pair 
(L3)(L6), (L2)(L6), (L3)(L5), (L1)(L6), (L4)(L6), (L4)(L5), (L2)(L5), (L1)(L5), (L3)(L4), (L5)(L6), (L1)(L3), (L1)(L4), (L2)(L3), (L2)(L4) and (L1)(L2) from left to right.}
      \label{mytt:fig6}
\end{figure}

Until now, for clarity of our explanation, we have restricted $x_{n}$
and $X_{n}$ to represent the expression (L1), and $y_{n}$ and $Y_{n}$
are for (L5), but the ratio $R(I_{t})$ defined by Eq.(\ref{mytt:eq9}) can be used for any
pair of expressions selected from the possible six expressions
(L1)$ \sim$(L6). Figs.\ref{mytt:fig6}(a) and (b) show the results of $R(I_{t})$
calculated for all the 15 possible pairs; (a) for the $[53:53]$ mode and
(b) for the $[53:64]$ mode. All the parametric conditions are the same
as those for Figs.\ref{mytt:fig5}(a) and (b).

Remarkable difference is that for the mode $[53:53]$ all the graphs for the
15 pairs are superposed into one, whereas for the mode $[53:64]$ each of the
15 transition graphs emerges at separate $I_{t}$ values with almost the
same slope which is much less steep than that of the transition curve
in Fig.\ref{mytt:fig6}(a). As far as the chaotic-orbit ratio $R(I_{t})$ is concerned,
it is independent of the  expression used in the computer
program for the mode $[53:53]$, whereas it is sensitively dependent on
the expression for the $[53:64]$ mode.

\section{Conclusions}
We first attracted attention of the computer users in scientific fields
to the fact that currently used FPU's have two computational modes to
handle the floating-point numbers in double precision(DP). We showed that
those two modes actually give rise to macroscopically observable
differences using the familiar logistic map as an example.

In papers reporting results of numerical calculations, what kind of
computer was used, programing language or codes themselves are seldom
included, let alone the computational mode examined in this report. One
of the most important objects of scientific papers, the present authors
believe, must be reproducibility of the results reported therein. Since
the fact that there are situations that the computational modes of FPU
in DP produce differences is made clear by this work,
the present authors propose that
in papers reporting numerical results, computational mode and expressions used should be
mentioned explicitly for further reproduction of the results. This proposal
does not at all urge every author to inspect his system to find out which
computational mode it utilizes. All he should do, we propose, is to include
a line that mentions what compiler was used on what machine in his numerical research.
Then, if the compiler is one of the popular ones, readers can see its computational
mode as indicated in section \ref{mytt:compmode}.

\section*{Appendix}
In this appendix,  we prove that for $a=4.0$ the expressions (L1)$\sim$(L6)
can be grouped into two groups as stated at the end of section 3.

We assume that for all expressions the exponents of the resulting floating-point
numbers become identical,  hence the two expressions are concluded to be
equal when the mantissae coincide.
We premise the following (A1),  (A2), and (A3) as formulas for our proof.
A floating-point number $x$ is defined as
\begin{align*}
      x = 2^{e}m(x)
\end{align*}
where $e$ is the exponent for the base 2 and $m(x)$ is the mantissa of $x$.

Multiplication by 4.0 does not alter the mantissa,  thus we have
\begin{align*}
        m(x) = m(4.0x).   	              \tag{A1}\label{a1}
\end{align*}
The mantissa of the product of two floating-point numbers $x$ and $y$ is given, 
in general, as the product of the mantissae of $x$ and $y$,  therefore we have
\begin{equation*}
	m(xy)=m(m(x)m(y)).                              \tag{A2}\label{a2}
\end{equation*}
Subtraction of two floating-point numbers $x$ and $y$ is done for the mantissae
after equalizing the two exponents.  If the differences of the two
exponents are the same,  the resulting mantissae are also the same.
Hence, we have
\begin{equation*}
m(4.0x-4.0y)=m(x-y),                              \tag{A3}\label{a3}
\end{equation*}

First we prove that (L1)=(L2)=(L3).
The expression (L1) is $(4.0x)(1.0-x)$, whose mantissa can be written
from (\ref{a2}) and (\ref{a1}) as
\begin{equation*}
\begin{array}{rl}
m((4.0x)(1.0-x)) & =m(m(4.0x)m(1.0-x))  \\
             & =m(m(x)m(1.0-x)). \\
\end{array}
\end{equation*}
The expression (L2) is $4.0(x(1.0-x))$,   whose mantissa can be written
from (\ref{a1}) and (\ref{a2}) as
\newlength{\len}
\begin{equation*}
\begin{array}{rl}
\settowidth{\len}{m(4.0(x(1.0-x))}
m(4.0(x(1.0-x))) &  =m(x(1.0-x)) \\
            & =m(m(x)m(1.0-x)). \\
\end{array}
\end{equation*}
Therefore,  we have proved that (L1)=(L2).
The expression (L3)  is $(4.0-4.0x)x$, whose mantissa can be written
from (\ref{a2}) and (\ref{a3}) as
\begin{equation*}
\begin{array}{rl}
m((4.0-4.0x)x) & =m(m(4.0-4.0x)m(x)) \\
           & =m(m(1.0-x)m(x))\\
           & =m(m(x)m(1.0-x)).\\
\end{array}
\end{equation*}
Since the order of multiplication does not matter,  we have
\begin{equation*}
m((4.0-4.0x)x)  =m(m(x)m(1.0-x)).\\
\end{equation*}
Therefore,  for $a=4.0$ we have proved that (L1)=(L2)=(L3).

Next, we prove that (L4)=(L5)=(L6). The expression (L4) is $4.0(x-xx)$,
whose mantissa can be written from (\ref{a1}) as
\begin{equation*}
\begin{array}{rl}
m(4.0(x-xx)) & =m(x-xx). \\
\end{array}
\end{equation*}
The expression (L6) is $4.0x - 4.0(xx)$, 
whose mantissa can be written from (\ref{a1}) as
\begin{equation*}
\begin{array}{rl}
m(4.0x-4.0(xx)) & =m(x-xx). \\
\end{array}
\end{equation*}
Hence,  we have proved (L4)=(L6) first.
The expression (L5) is $4.0x - (4.0x)x$, 
whose first term is identical with that of (L6).
Therefore, if the second terms of (L5) and (L6) are identical, 
we can conclude that (L5)=(L6).
The mantissa of the second term of (L5) can be written from (\ref{a2}) 
and  (\ref{a1})  as
\begin{equation*}
\begin{array}{rl}
m((4.0x)x) & =m(m(4.0x)m(x)) \\
             & =m(m(x)m(x))).  \\
\end{array}
\end{equation*}
The mantissa of the second term of (L6) can be written from (\ref{a1}) 
and  (\ref{a2}) as
\begin{equation*}
\begin{array}{rl}
m(4.0(xx)) & =m(xx) \\
           & =m(m(x)m(x)).  \\
\end{array}
\end{equation*}
Hence we see that the second terms of (L5) and (L6) are identical.
We have proved that (L4)=(L5)=(L6).

\bibliographystyle{ws-ijbc}
\bibliography{sample}

\end{document}